\documentclass[%
reprint,
groupedaddress,
showpacs,preprintnumbers,
nofootinbib,
 amsmath,amssymb,
 aps,
prstab,
floatfix,
longbibliography
]{revtex4-1}

\usepackage{graphicx}
\usepackage{dcolumn}
\usepackage{bm}
\usepackage{ifthen}

\begin{document}

\preprint{FERMILAB-PUB-12-640-AD}

\title{The Fermilab Main Injector: high intensity operation and beam loss control}

\author{Bruce C. Brown}
\author{Philip Adamson}
\author{David Capista}
\author{Weiren Chou}
\author{Ioanis Kourbanis}
\author{Denton K Morris}
\author{Kiyomi Seiya}
\author{Guan Hong Wu}
\author{Ming-Jen Yang}
\affiliation{Accelerator Division, Fermi National Accelerator Laboratory,
Batavia, Il, 60510, USA}
\email[]{bcbrown@fnal.gov}
\thanks{ Operated by Fermi Research Alliance, LLC under Contract No. DE-AC02-07CH11359 with the United States Department of Energy}

\date{06/19/2013}

\begin{abstract}

From 2005 through 2012, the Fermilab Main Injector provided intense
beams of 120 GeV protons to produce neutrino beams and
antiprotons. Hardware improvements in conjunction with improved
diagnostics allowed the system to reach sustained operation at ~400 kW
beam power. Transmission was very high except for beam lost at or near
the 8 GeV injection energy where 95\% beam transmission results in
about 1.5 kW of beam loss. By minimizing and localizing loss, residual
radiation levels fell while beam power was doubled. Lost beam was
directed to either the collimation system or to the beam
abort. Critical apertures were increased while improved
instrumentation allowed optimal use of available apertures. We will
summarize the improvements required to achieve high intensity, the
impact of various loss control tools and the status and trends in
residual radiation in the Main Injector.

\end{abstract}

\pacs{29.20.dk, 29.27.Fh, 41.85.Ja, 41.85.Si}
\maketitle


\newlength{\FigResize}

\section{\label{PROTONS}Protons to Produce Neutrinos and Antiprotons}

On April 30, 2012, the Fermilab accelerator complex began an extended
shutdown for facility upgrades.  This followed seven months after the end of operation for
the Tevatron on September 30, 2011 with the accompanying end of
antiproton source operation.  For the Fermilab Main Injector, this
marked $13\frac{1}{2}$ years of commissioning and operation in
successively higher intensity operation modes.  As the physics program
requirements demanded more beam power, limitations in the intensity
and beam quality from the Fermilab Booster were overcome by using slip
stacking injection~\cite{Seiya:PAC2007}.  This was implemented first
for antiproton (pbar) production and later for neutrino production as
well.  Once this concept was proven, required upgrades to the Linac,
Booster and Main Injector to support high intensity operation were
identified and a Proton Plan~\cite{BeamsDoc1441,ProtonPlan} organized
to implement them.  As intensities increased, a program of monitoring
and mitigating losses and residual radiation has controlled the
radiation exposure for personnel involved in maintenance and upgrade
activities.

\begin{figure}[b]
   \includegraphics*[width=\columnwidth]{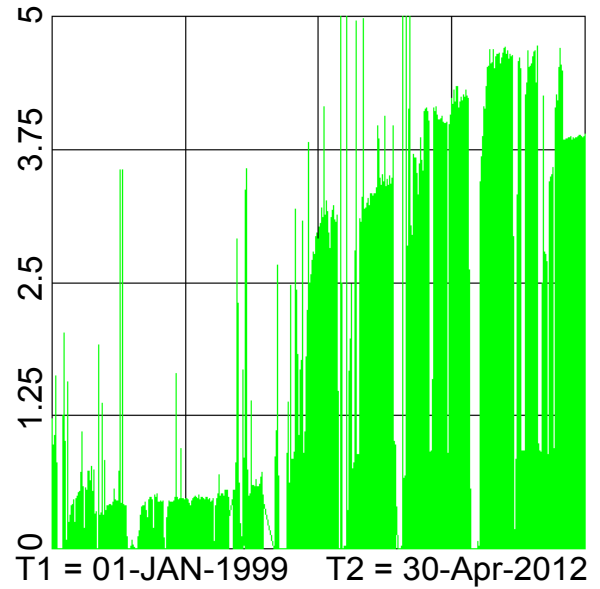}
   \caption{\label{Fig_IntHist}Sampled Intensity per cycle from 
    January 1999 through April 2012 with full scale of 
    $5\times10^{13}$ protons per pulse (50 Tp  per pulse).
    Vertical grid lines are at 2 May 2002, 31 August 2005, and
    30 December 2008,   }
\end{figure}

Figure~\ref{Fig_IntHist} illustrates this intensity increase using the
number of protons per cycle on a periodic sample of the acceleration
cycles.  An injection from the Booster is termed a `batch' with
typical intensity of $4$--$5\times10^{12}$ protons and up to 84 rf
buckets of beam.  Machine commissioning was followed by multibatch
operation for a Tevatron fixed target run.  In 2001, this transitioned
to a Tevatron collider run which utilized a single batch from the
Booster for pbar production. Slip stacking injection of two Booster
batches for pbar production became operational in December 2004.  In
May 2005, the NuMI (Neutrinos at the Main Injector) beamline for
neutrino production began operation which required each acceleration
cycle to provide 5 batches to be sent to the NuMI target.  Mixed Mode
slip stacking injection became operational in August 2005 with 5
batches to be sent to NuMI plus a double (slip stacked) batch for pbar
production (5+2 cycle).  Eleven batch Mixed Mode slip stacking (9+2
cycle) which provides four double batches for increased NuMI beam was
commissioned in January 2008 at the same time as was the Main Injector
collimation system~\cite{Brown:WE6RFP025}.  At that point, intensity was
limited by losses in both the Main Injector and the Booster.
Collimation, along with improved Booster beam quality, controlled
activation and permitted Main Injector intensity per cycle to
increase.

Several other features of the Fermilab High Energy Physics (HEP)
program are apparent in Fig.~\ref{Fig_IntHist}.  Facility upgrades are
accomplished using shutdown periods of several weeks.  Periods of
reduced intensity mark Tevatron failures or the time required to
repair or replace the NuMI horn or target.  When pbar production
ended, intensity ceilings were needed to limit neutrino target thermal
shock.  Accelerating cycles with 9 batches, including three which were
slip stacked, provided the required intensity.  The reduced per pulse
intensity from October 2011 through April 2012 reflects this
limitation.  This figure reports measurements from older
instrumentation or data recording for which improved systems were
available by 2007 and spikes above the trend are typically due to
instrumentation or data recording errors.

\begingroup
\squeezetable
\begin{table}[tb]
  \caption{\label{Table:BeamProp}Main Injector Properties Including Typical 
Beam Properties for High Intensity 120 GeV Operation}
\begin{ruledtabular}
\begin{tabular}{lccc}
Lattice Properties  \\
Measured Circumference & 3319.4151  & m \\
Courant-Snyder Amplitude $\beta_{max}$ & 57 & m \\
Courant-Snyder Amplitude $\beta_{min}$ & 10 & m \\
Maximum Dispersion Function & 1.9 & m \\
Transverse Admittance    & $>40\pi$   & mm-mr \\
Longitudinal Admittance  & $>0.5$   & eVs \\
Nominal Horizontal Tune & 26.425 &\\
Nominal Vertical Tune   & 25.415 &\\
Natural Horizontal Chromaticity & -33.6 & \\
Natural Vertical Chromaticity & -33.9 & \\
Transition $\gamma$ & 21.8 & \\
\\
RF Properties \\
Booster Harmonic Number& 84 &         \\
Main Injector Harmonic Number & 588    \\
RF Frequency (Injection)& 52.811& MHz\\
RF Frequency (Extraction)&53.104& MHz\\
RF Accelerating Cavities &18&\\
Peak RF Voltage          & 4 & MV\\
Maximum Acceleration Rate & 204 & GeV/s&\\
\\
Nominal Injected Beam Properties    \\
Kinetic Energy  &	8   &	GeV \\
Transverse Emittance (95\%) &  15$\pi$& mm-mr \\
Longitudinal Emittance per bunch (95\%) & 0.08&   eVs\\
Momentum Spread ($\delta p$ 95\%)    &  	8   &  	MeV/c \\
Bunches transferred per Booster cycle& 81 &\\
\\
Mixed Mode Operation - pbar Beam      \\
Booster Beam Intensity& 4.3$\times 10^{12}$& protons/Batch\\
Number of Batches     &    2  &             \\         
Transmission Efficiency&93\%  &             \\
Extracted Beam Intensity& 8$\times 10^{12}$&	protons/cycle\\
\\	
Mixed Mode Operation - NuMI Beam          \\
Booster Beam Intensity & 4$\times 10^{12}$ & protons/Batch \\
Number of Batches      & 9    &              \\
Transmission Efficiency& 95\% &              \\
Extracted Beam Intensity& 34$\times 10^{12}$& protons/cycle 	\\
\\
Slip Stack Frequency Difference&1430& 	Hz \\
Slip Stack Interval     &  5/15     &	seconds\\
\\
Typical Extracted Intensity(NuMI+pbar)&42$\times 10^{12}$&protons/cycle\\
Record Extracted Intensity &	46.3$\times 10^{12}$&protons/cycle 	\\
Main Injector Cycle Time (Mixed Mode)  & 2.2 &seconds \\
Beam Power (typical)       & 380 & kWatts \\
Beam Power (record for 1 hour)&  400& 	kWatts\\
\end{tabular}
\end{ruledtabular}
\end{table}
\endgroup

\section{\label{MIIntro}Introduction to the Main Injector}

The Fermilab Main Injector Project~\cite{FMITDH} was created to enhance
the physics capabilities of the Tevatron collider and to provide beams
of 120 GeV protons for test beams and fixed-target particle physics
experiments.  The initial goal for high intensity operation was
3$\times$10$^{13}$ protons per pulse at 120 GeV.  Construction of the
Main Injector began in June of 1992 with commissioning beginning in
September 1998.  Basic properties of the Main Injector and important
features for high intensity operation are summarized in Table
\ref{Table:BeamProp}.  With approval of the NuMI neutrino beam and the
MINOS experiment, high intensity operation became the focus for
upgrades.

In addition to providing 120 GeV protons at high intensity, the Main
Injector was required to: accelerate high quality beams of protons and
antiprotons to 150 GeV for injection to the Tevatron, to supply 120
GeV protons to fixed target experiments and test beams, and to
transfer antiproton beams at 8 GeV to and from the permanent magnet
Recycler Ring in the same tunnel and from the Antiproton Source.
Transfers of 8 GeV protons were required for tuning the transfer lines
and the Accumulator and Debuncher (Antiproton Source).

This document is devoted to the high intensity operation~\cite{* [{The
loss control section is expanded from }] [{}] Brown:HB2012}.  We will
describe instrumentation improvements and the residual radiation
monitoring program.  An overview of slip stacking injection and rf
modifications required to achieve high per pulse intensities will be
provided.  Dampers needed to control instabilities, aperture
improvements which reduce losses, and collimation systems to localize
the remaining beam loss will be discussed.  We will describe minor
problems which were resolved so that we achieved low residual
radiation nearly everywhere in the Main Injector enclosure. The beam
power and proton intensities delivered will be summarized.

\begin{figure*}[htb]
   \includegraphics*[width=\textwidth]{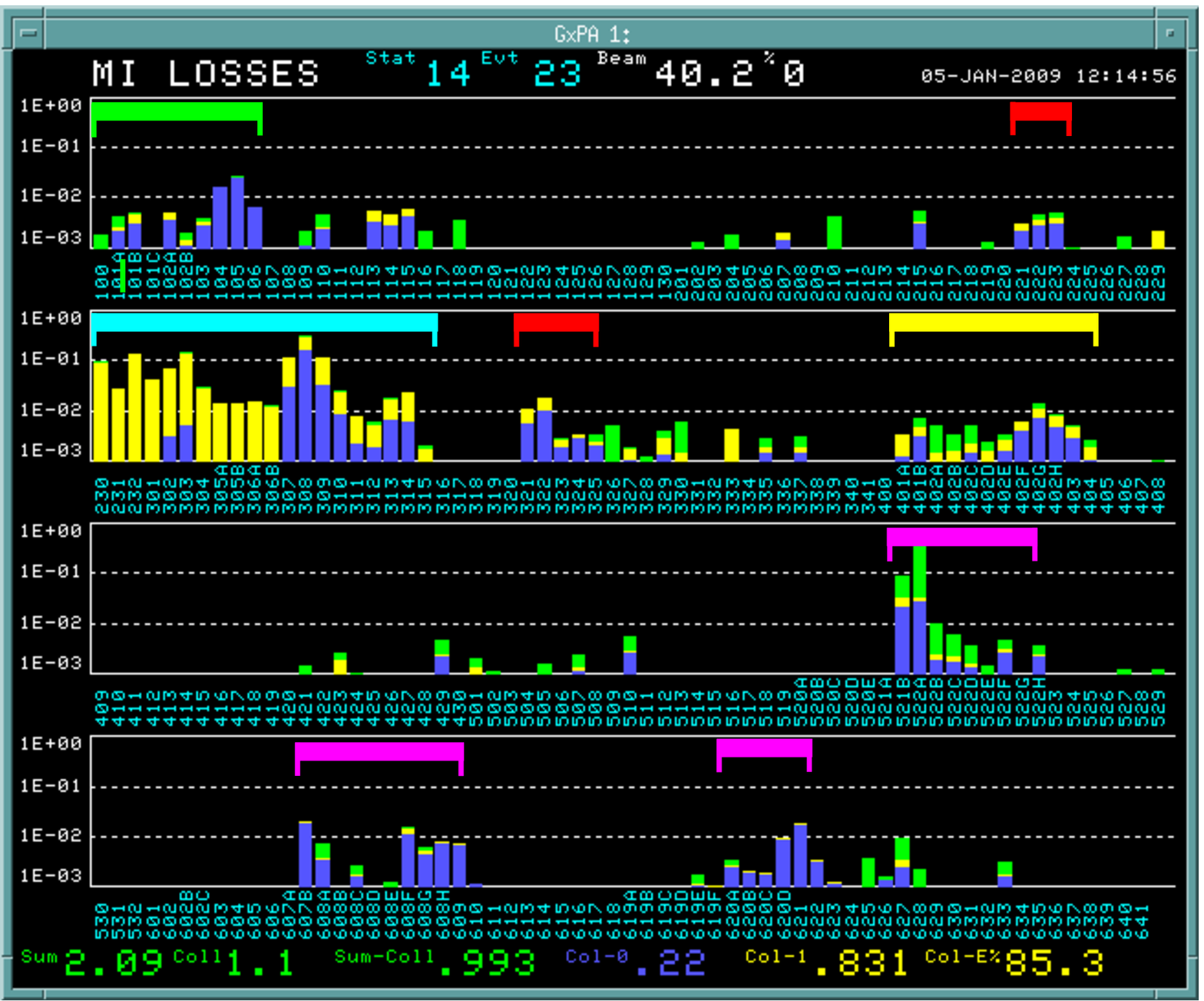}
   \includegraphics*[width=\textwidth]{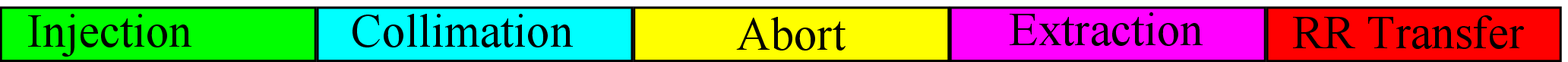}
   \caption{    \label{BLM2009}
Beam Loss Monitor Display from a Main Injector Cycle in January 2009 at
an intensity of $4.4\times10^{13}$ protons.  The three decade
logarithmic vertical scale in Rads/pulse ranges from 1 milliRad to 1
Rad.  Values shown are integrated loss at each BLM in the ring at end
of cycle in green, overlayed in yellow by loss integral after 1.5\%
acceleration, then overlayed in blue by the loss integral at the end of
injection.  In typical operation, most of the green loss results from
the extraction process at the very end of the cycle.  Functional areas
of the ring are marked by colored braces which are identified
below the display.}
    \end{figure*}

\section{\label{UpgradeIMC}Upgrades for High Intensity: 
Instrumentation, Monitoring and Control}

The instrumentation for commissioning the Main Injector used data
acquisition and electronic systems developed for the Fermilab Main
Ring in the 1980's.  By 2006 new systems were commissioned.  The new
beam position monitor (BPM) system employs digital signal receivers
for signal conversion.  Enhanced flexibility as well as improved
resolution for position measurement are
available~\cite{Banerjee:2006zj}.  For the 250 ionization chamber beam
loss monitors (BLM's)~\cite{Shafer-TeV-BPM-BLM}, a new digitization
and data recording system~\cite{Baumbaugh:2011aaa} provides
flexibility for studies and much enhanced monitoring capability.  A
more sophisticated data collection system from the existing beam
current monitors was developed using a stand alone micro-processor
(BEAMS front-end~\cite{* [{The beam current monitor software described
in}] [{}] 1748-0221-6-11-T11004}).Together these new instruments
allowed a more systematic study of the machine and improved displays
of routine operation.

New control console programs were developed to employ the BLM
system~\cite{BeamsDoc3299v2}.  For studies, a flexible system to set
data collection times provided details about loss mechanisms by
allowing time correlated measurements on all loss monitors. Beam loss
displays were particularly significant for improving the overall loss
pattern by emphasizing high losses while disclosing lesser beam loss
locations which had previously gone unobserved.  Fig.~\ref{BLM2009}
shows beam losses for operations in January 2009.  This display
occupies a prominent place in the Fermilab Accelerator Control Room.

\begin{figure*}[tb]
   \includegraphics*[width=\textwidth]{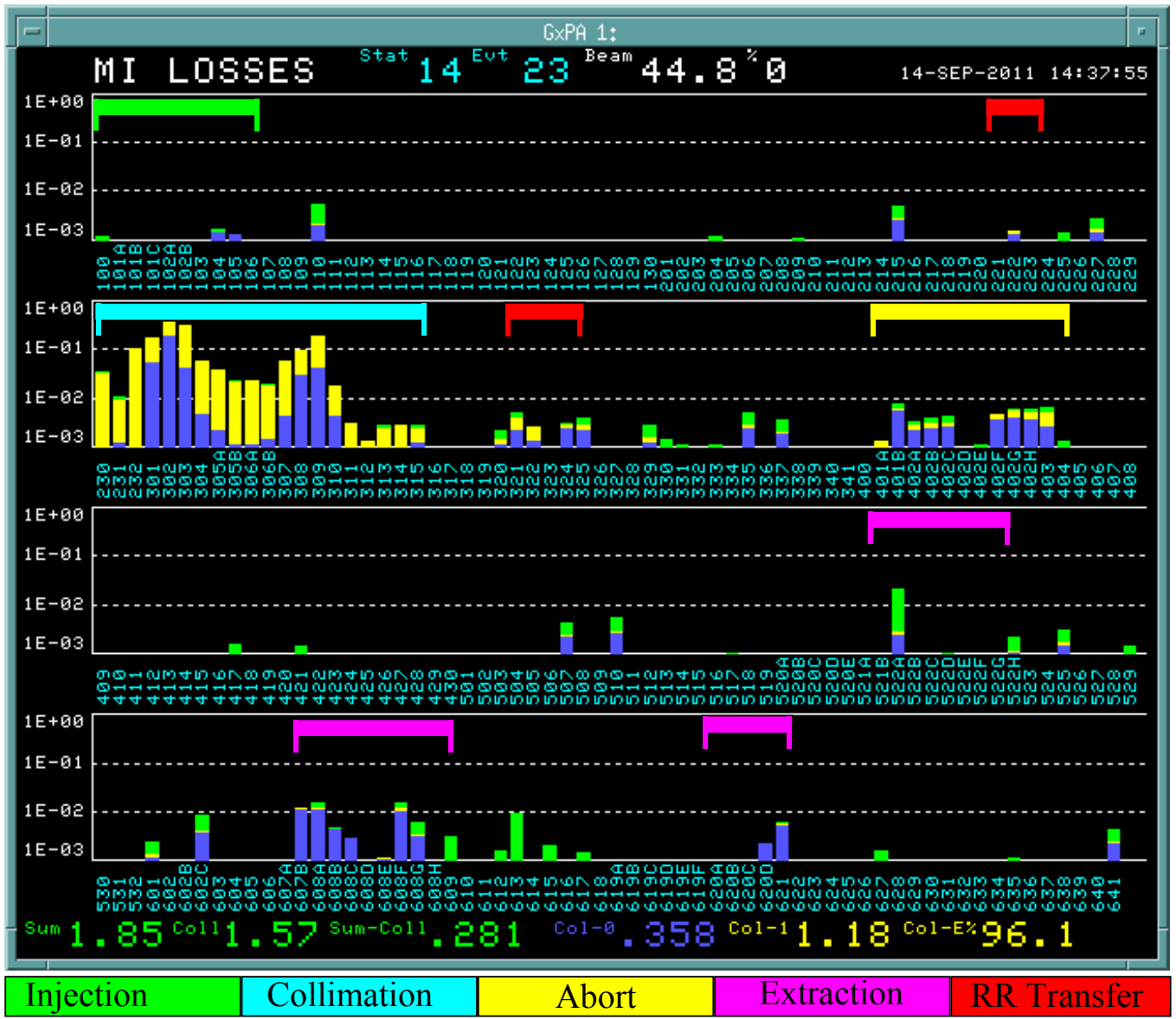}
   \includegraphics*[width=\textwidth]{LossPlotLabels.eps}
   \caption{\label{BLM2011} Beam Loss Monitor Display from Main
Injector Cycle in September 2011 at an intensity of $4\times10^{13}$
protons prior to the end of pbar production.  Loss monitors are shown
sequentially around the Main Injector.  Compare to Fig.~\ref{BLM2009}
above. Some remaining BLM signals are due to pedestal offset, not 
beam loss. }
    \end{figure*}

Injection period loss (blue) are seen all around the ring.  Those in
the injection region (green brace) are created by circulating beam in
the injection gap.  Beam lost during early acceleration (yellow) are
due to unaccelerated (uncaptured) beam (as described in
Section~\ref{SSLosses}).  By this time (2009), the collimation
system (see section~\ref{LossContColl}) was beginning to localize
these losses at the collimator region (cyan brace) but they are still
seen in many other locations.  Further efforts were required.  End of 
cycle losses (green), when not overlayed by earlier loss integrals, are
typically from the extraction process.  They are apparent at both 
Recycler transfer regions (red brace), at the abort  location (yellow brace),
and the  high energy transfer locations (purple brace).  Other locations
with no special lattice function also show loss.  Fig.~\ref{BLM2011},  from
2011,  illustrates  the  progress  documented in  this  paper.   These
results  are  discussed  further  in  Section~\ref{BLMDisplay}.   Many
losses were reduced by only employing proper orbit correction.

Preparations for the high intensity operation for neutrino production
included a program to identify residual radiation issues in the Main
Injector tunnel.  Exploratory residual radiation measurements in 2004
and 2005 monitored more than 100 locations with more than 20
milliRad/hr residual radiation on contact.  By October 2005, a program
using a sensitive meter to monitor 127 (later expanded to 142)
bar-coded locations was initiated~\cite{BeamsDoc3523v1}. Loss issues at
beam transfer regions were monitored and some unexpected loss patterns
were identified and explored (see Section\ref{PipeAlignment}).  The
need for loss localization using collimators was documented.  See
Section~\ref{RRMonitor} and Fig.~\ref{RR100-400-100} for some results
from these measurements.

\begin{figure*}[tb]
\includegraphics*[width=0.45\textwidth]{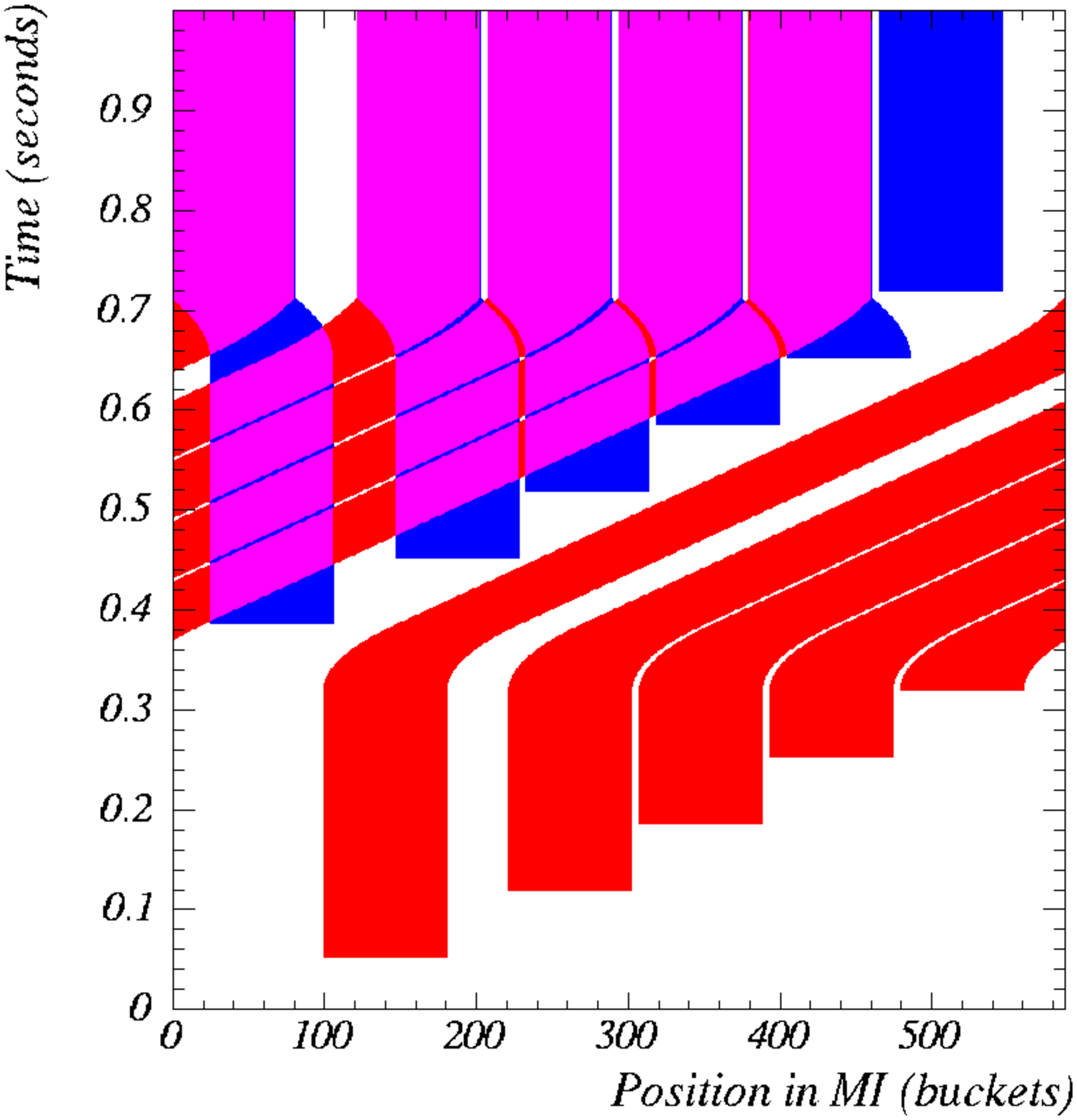}
\includegraphics*[width=0.45\textwidth]{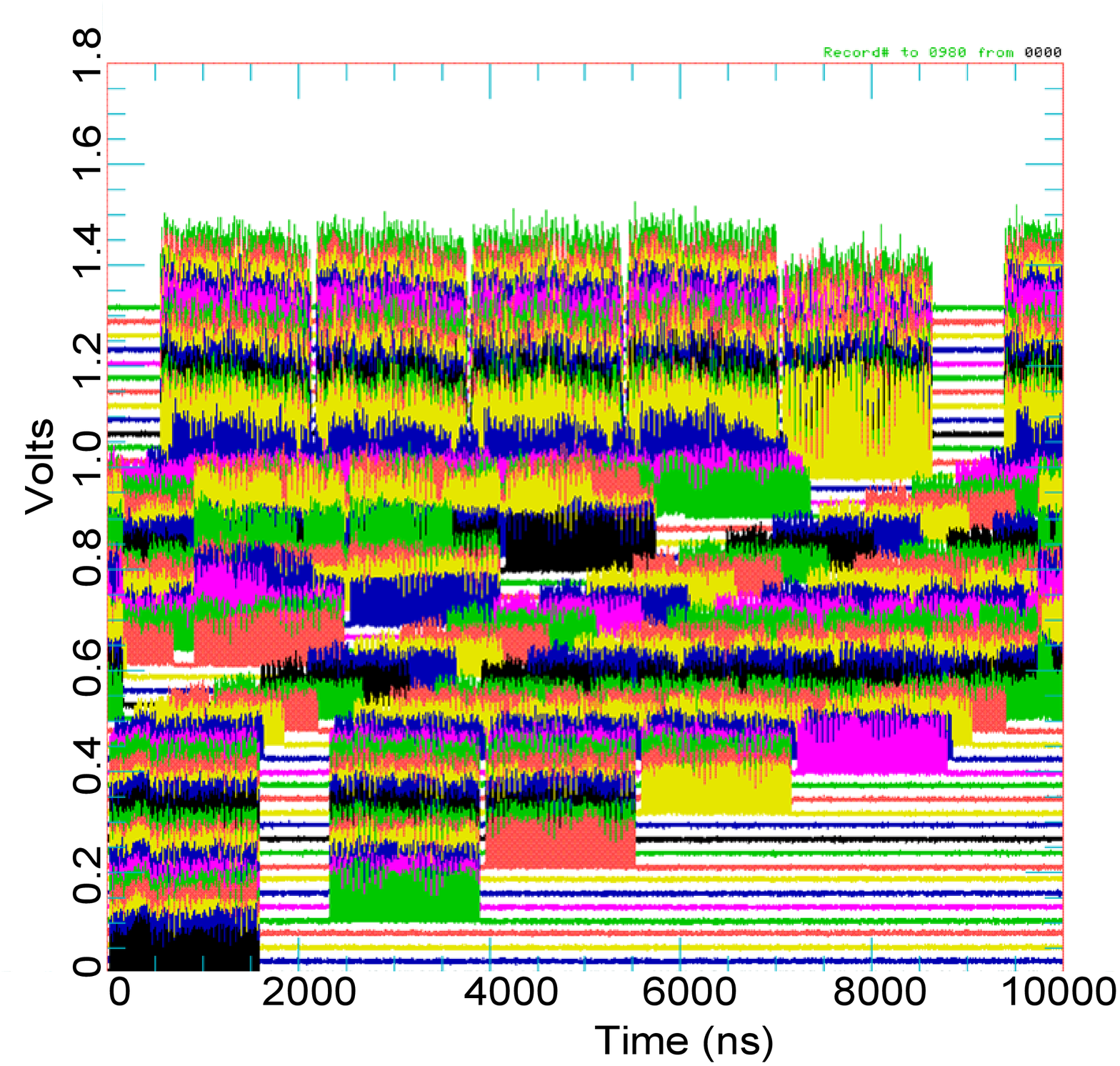}
\caption{Left panel shows logic of slip stacking (see
Section~\ref{SSMechanism}) Five (red) Booster batches are injected
into buckets of the first rf system.  The frequency is reduced to
decelerate them and the next five (blue) batches are injected.  When
the bunches in the red and blue batches are aligned, the 1 MV rf
system captures them.  Then the final blue batch is injected.  Right
panel shows Wall Current Monitor signal during injection for slip
stacking operation using 11 Booster batches. Horizontal axis shows
time for (nearly) one Main Injector revolution.  Main Injector
revolution time is 11134 ns at injection.  Vertical axis shows bunch
intensity with later turns offset vertically.  Four double batches
followed by one single batch for neutrino production are phased for
acceleration at the top of the figure. The first injected batch for
pbar production begins at the lower left, slips to the right and is
joined by the second pbar injection, arriving at acceleration phase
before the top of the figure where it is at the far right.  The final
neutrino batch arrives following the recapture to the left of the gap
for pbar extraction.  This panel is from
Reference~\cite{Seiya:PAC2007}.  With this mode, Main Injector
intensities of up to $4.6\times10^{13}$ protons per cycle are
achieved.  \label{11BatchSlipStack}}
\end{figure*}

\subsection{Injection Line Collimation}

The exploratory residual radiation monitoring program included detailed
studies of the radiation pattern which suggested that beam halo was
greatly increasing the number of radioactive locations.  Collimation
of the Booster beam in the transfer line was an obvious option.  In
order to collimate beam in a transfer line, in both horizontal and
vertical planes it is necessary to have collimation edges on two sides
of the beam and at two locations.  This was accomplished in the
Fermilab Booster to Main Injector transfer line (MI8 Line) with the
corners of four rectangular apertures using pairs of collimators at
two locations separated by $90^o$ phase advance.  This collimation
system~\cite{PAC07_CSFBMITL} was installed in 2006 and has operated to
scrape beam edges beyond about 99\% of the beam.  Beam orbit drift
would cause fluctuations in the transmitted beam by asymmetric
collimation.  This was controlled by an auto-tune system with frequent
beam position measurements to determine new trim magnet settings
resulting in stability at the $\sim0.1$~-~$0.2$ mm level.

\subsection{\label{InstabControl}Instability Control for High Intensity Operation}

Commissioning of the Main Injector achieved a goal of accelerating
$2\times10^{13}$ protons per cycle.  However, this required use of
large negative chromaticity to control the resistive wall instability.
The resulting beam lifetime at injection energy resulted in beam
losses of ~10\%.  This loss was alleviated by using the transverse
mode damping provided by a digital damper
system~\cite{Adamson:2005bw}.  Very high transmission is achieved with
near-zero chromaticity when using the digital dampers.

Longitudinal damping by this system improved the longitudinal
emittance by damping injection oscillations from the Booster and by
avoiding coupled bunch instabilities in the Main Injector seeded by
the Booster oscillations.  This smaller emittance is important in
achieving the shortest possible bunch length for efficient pbar
production.  Longitudinal control was also important for Tevatron
injection.  The longitudinal dampers improved the efficiency of slip
stacking (see Section~\ref{SlipStacking} below) by ~1\% permitting
higher intensities and also modestly improved the ability to control
losses at Main Injector transition.

\section{\label{SlipStacking}Slip Stacking to Achieve High Intensity}

\begin{figure}[b]
   \includegraphics*[width=65mm]{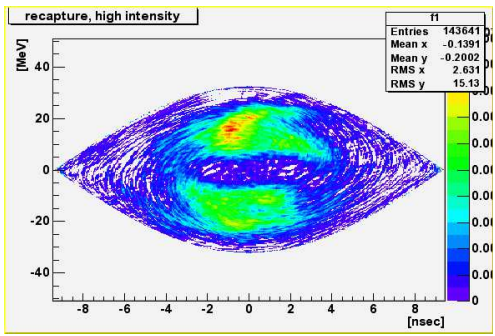}
   \caption{\label{RecaptureTomograph}Tomograph showing momentum
offset against time for a recaptured
bucket of slip stacked beam as reconstructed from resistive wall monitor
signal after creation of 1 MV rf capture bucket.  This figure is from
Reference~\cite{Seiya:2006zz}}
\end{figure}

The 40 year old Fermilab Booster provides the 8 GeV beam injected into
the Main Injector.  The Main Injector circumference is 7 times that of
the Booster\footnote{The Booster employs harmonic number 84.  For high
intensity operation, the Main Injector employs harmonics number 588
using a 53 MHz rf system.  A 2.5 MHz system was employed for some pbar
operations.}  but the need for clean transfers limits operation to 6
Booster batches, leaving time for the rise and fall of the fields in
the transfer kicker magnets (kicker gaps).  Following the 400 MeV
upgrade of the Linear Accelerator~\cite{Junck:1994gy} which injects
into Booster, it was found that intensities up to $5.5\times10^{12}$
protons per pulse could be accelerated.  Losses in Booster and output
beam quality limited useful operation to $5\times10^{12}$ and beam
quality was improved by operation at lower
intensities~\cite{Webber:2000yh}. Injection of 6 batches at
$5\times10^{12}$ protons will only provide $3\times10^{13}$ protons
per Main Injector cycle\footnote{A cycle delivering beam to pbar, then
NuMI, required gaps for both the rise and fall of the kicker pulse
which delivered beam to the pbar target.}. The Fermilab Antiproton
Source employed Booster-length Debuncher and Accumulator Rings so
increased pbar production depended on higher Main Injector intensity
concentrated in one batch length.  The capabilities of the neutrino
program was limited without enhanced Main Injector beam power.

\subsection{\label{SSMechanism}Slip Stacking Mechanism}

\ifthenelse{\lengthtest{\columnwidth>100mm}}{\setlength{\FigResize}{100mm}}{\setlength{\FigResize}{\columnwidth}}

\begin{figure}[tb]
\includegraphics[width=\FigResize]{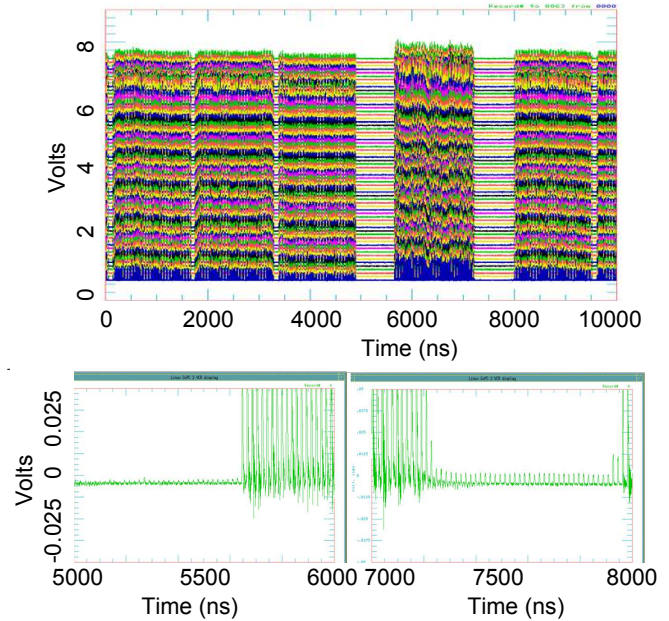}
\caption{Wall Current Monitor plot for slip stacked beam during
development of (5 + 2) slip stacking.  For the upper panel, the
horizontal axis shows time for (nearly) one Main Injector revolution.
Vertical axis shows bunch intensity after recapture with selected
turns offset vertically.  The lower panel shows portions of one turn,
just before extraction, expanded to show the beam captured in the gaps
for the rising (left) and falling (right) edges of pbar extraction
kicker pulse. This figure is from Reference~\cite{Seiya:PAC2007}.
\label{SlipStackWCM}}
\end{figure}

Slip stacking injection allows higher intensity by employing the
momentum aperture of the Main Injector to circulate pairs of Booster
batches at different momentum, allowing them to slip into alignment
for recapture by the accelerating rf waveform.  Double batches for
pbar production began by using one batch delivered to the central
orbit using bucket-to-bucket transfer into a 100 kV rf bucket at the
nominal rf frequency in the Main Injector.  After decelerating it to a
lower momentum (by lowering the rf frequency), a second batch is
transferred bucket-to-bucket into an adjacent longitudinal position
using a separate rf system at the injection frequency.  The buckets in
these rf systems slip with respect to each other.  After the 10th
injection, both rf frequencies are increased so that they are
symmetric above and below the central orbit frequency.  At the moment
when the bunches are aligned, the 1 MV acceleration rf system captures
both batches in a single larger bucket. For multi-batch slip stacking,
five batches are injected before the deceleration, then five more are
added before recapture. The remaining location is filled with a single
batch. Fig.~\ref{11BatchSlipStack} illustrates this injection process.

Following initial demonstrations of slip stacking, measurements and
simulations were carried out~\cite{Seiya:PAC2005} which revealed
required upgrades and limitations.  Beam loading compensation for the
Main Injector rf was required for adequate capture in the lower
voltage slip stacking buckets~\cite{Dey:2005jw}.  Bunch rotation in
the Booster to reduce $\delta p/p$ was also required to match these
buckets for bucket-to-bucket transfer.  These developments required
simulation with ESME~\cite{ESME} and other longitudinal space
simulations as well as measurements of Booster and Main Injector beam
properties~\cite{Seiya:PAC2007}.  Measurements of the recaptured beam
bunch is shown in Fig.~\ref{RecaptureTomograph}. The simulations,
along with measured beam properties, documented the requirements for
Booster beam properties (see Table~\ref{Table:BeamProp}) but also
showed that beam loss was expected.

\subsection{\label{SSLosses}Losses from Slip Stacking}

For sufficiently small emittances, capture efficiencies with slip
stacking can be very high.  Since the damper system acts on single 53
MHz bunches, it is unsuitable for controlling instabilities during
the slipping process.  As a result, when bunches are slipping, the
instabilities must be controlled by jumping the chromaticity to a
large negative value (-20) which results in some beam loss.

Matching the slipping time to the Booster cycle sets the frequency
separation required for the counter-slipping bunches.  This determines
the usable bucket area and sets the longitudinal admittance for slip
stacking injection.  The momentum acceptance of the Main Injector
accepts these two bucket streams, but for the desired intensity, the
Booster emittance is a bit too large.  This results in various loss
issues.  Beam which is outside of the slip stacking buckets can move
in longitudinal phase such that either:
\begin{enumerate}
\item It is re-captured in an extraction kicker gap.
\item It has drifted into an injection kicker gap.
\item At re-capture time it is outside of the 1 MV bucket and
will not be accelerated.
\end{enumerate}
In addition, when the beam is subjected to high negative chromaticity,
the beam lifetime is reduced by transverse loss mechanisms.  
Control measures for all these losses are now
described\footnote{Simulation and measurement showed that the stable
bucket area for the first slip stacking rf system was reduced when the
second (higher frequency) system was turned on.  For operations, the
voltage from the second system was not turned on until required for
the 5th through 10th injections.}.

\subsection{\label{ExtGapLoss}Control of Extraction Gap Loss with Anti-Damping}

Using the wall current monitor, we illustrate the first of these
problems with Fig.~\ref{SlipStackWCM}. In the upper figure we see the
5 NuMI and 1 pbar batches.  The pbar production batch is now just to
the right of center.  We see gaps for the rise (left) and fall (right)
of the pbar extraction kicker.  These gaps are expanded in the lower
panel showing beam which was re-captured and accelerated.  Details of
the extraction gap beam varies primarily because of variation in the
Booster beam momentum distribution.  At 8 GeV, loss of ~1-2\% of the
protons is manageable. But after acceleration, losing the same protons
at 120 GeV creates more activation near extraction devices which may
require maintenance. Removal of these losses is essential.

At low energies, beam bunches in the gaps can be anti-damped to
achieve removal.  Initially this was accomplished by driving the
vertical bunch-by-bunch dampers open-loop at the fractional machine
tune~\cite{Adamson:2005bw}.  Later, a vertical kicker was installed
near the injection region.  Anti-damping with this device was
rudimentary but effective.  A bunch by bunch ~5 microradians vertical
kick at a $\beta_v = \,\, \sim50$~m, is turned on or off in accordance
with an assumed tune near the operating fractional beam tune of
$\sim$0.42.  The vertical emittance grows until the protons strike the
aperture limit. The assumed tune is programmed in steps of 0.01
applied for 1000 turns.  The tunes and steps are modified to optimize
beam removal.  Much of the proton loss strikes the secondary
collimators described below. Loss at the MI522 Lambertson (LAM522) has
been significant (see Fig.~\ref{RRhistory}).  During slip stacking
when there was beam captured in the extraction kicker gaps, losses
were capable of exceeding operating loss limits.  Loss control by
antidamping in these extraction gaps combined with the collimator and
gap clearing kicker systems resulted in extended periods with no
measurable loss at LAM522.

\subsection{\label{InjGapLoss}Control of Injection Gap Loss with Gap Clearing Kickers}

The two rf systems used for slip stacking define separate stable
buckets for maintaining bunched beams.  Beam outside of those stable
buckets will drift longitudinally on the slipping orbits. As discussed
in subsection~\ref{ExtGapLoss}, beam which is captured in the
extraction gaps will create losses.  The injection process transfers
beam in a series of buckets into the Main Injector using a Lambertson
magnet and vertical kicker (K103).  Any circulating beam in the ring
which passes through the kicker during the injection pulse will be
deflected and will strike magnets downstream of the injection kicker
in MI104 - MI106 (see Figures~\ref{BLM2009} ~and ~\ref{RRhistory}).
This beam is typically unbunched and is a problem as soon as 1/15
second after it was injected, making anti-damping ineffective.  The
solution was a system of gap clearing kickers
(GCK)~\cite{Kourbanis:WEP205} which are fired to clear the injection
gap just prior to the next injection, sending this beam to the Main
Injector Abort dump.  Prior to commissioning the GCK in 2010, the
residual radiation build-up in the injection region was minimized by
observing limits on the beam loss which was monitored with BLM and
residual radiation measurements.

\subsection{\label{ColOverview}Collimation Overview}

\begin{figure}[b]
   \includegraphics*[width=\columnwidth]{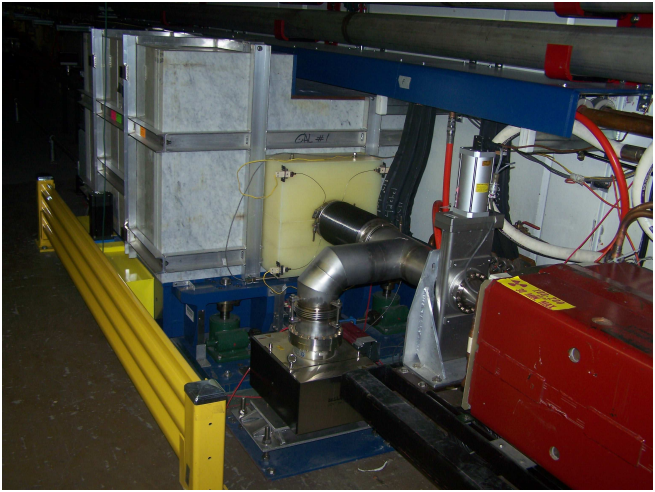}
   \caption{\label{20TCol}One of four secondary collimators which
employ a thick stainless steel vacuum chamber surrounded by a steel
absorber to contain the shower particles. Radiation shielding for
personnel is provided by 12 cm of marble placed on the top, ends, and
aisle side of the collimator.  At the upstream end a polyethylene
block reduces neutron flux to upstream magnet coils. At the downstream
end, masks are placed to absorb small angle outscattered particles.}
\end{figure}

\begin{figure}[t]
   \includegraphics*[width=\columnwidth]{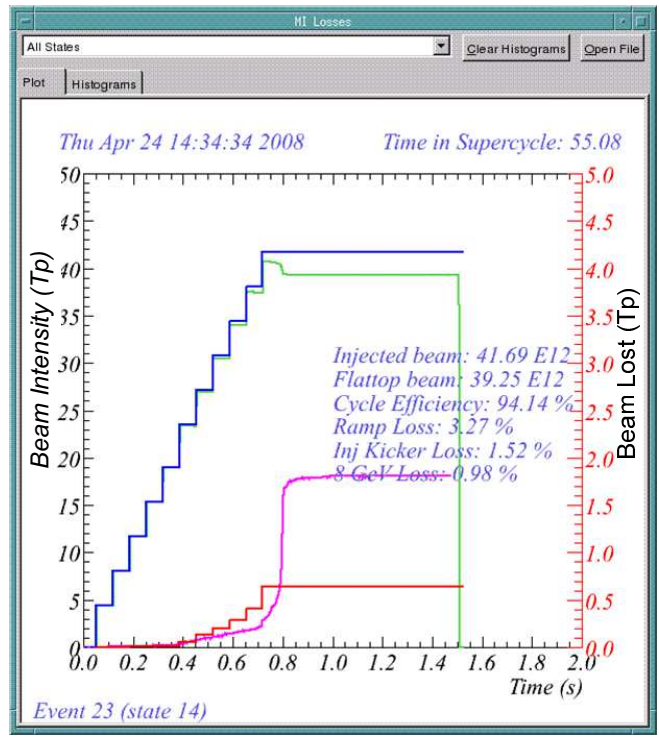}
   \caption{\label{AccelBeam}Typical Beam operation for pbar plus NuMI
beam production shows the beam intensity {\em vs.} time in the
acceleration cycle.  Acceleration begins at 0.755 seconds.
Intensities shown are blue (sum of injected beam), green (circulating
beam in Main Injector), red (loss from injection process), and magenta
(total loss in ring).}
\end{figure}

An additional loss due to slip stacking is from beam which is not
captured in the 1~MV rf buckets and thus not accelerated.  This beam
will follow the momentum offset orbit to lower momentum until the
machine aperture is reached. The Main Injector collimation
system~\cite{Brown:WGC11} localizes this loss to limit personnel
exposure.  It employs a primary-secondary collimator system which
defines the momentum aperture with a 0.25 mm tungsten primary
collimator located in a cell (MI230) with normal high dispersion which
is just upstream of the dispersion suppressor cells leading to the
MI300 straight section.  The vertical edge of this collimator is
positioned radially inside of the circulating beam to define the
momentum aperture.  As the beam reaches this aperture it is scattered.
Four 20-ton secondary collimators, such as the one shown in
Fig.~\ref{20TCol}, placed at appropriate phase advance downstream,
absorb ~80\% of the lost beam power with the rest going to nearby
devices in the collimation region. Particles which do not experience a
sufficient initial scatter may strike the primary collimator two or
three times before being lost from the circulating beam.  The loss
pattern is distinctive due to the narrow time structure of the
unaccelerated 8 GeV beam moving to the low momentum dispersion
orbit. Using the time structure as a diagnostic, examination of the
ring loss pattern shows that 99\% of the radiation from this beam loss
is captured in the collimation region~\cite{Brown:WE6RFP025}.  At
Booster intensity of $4.3 \times 10^{12}$ protons per Booster cycle,
the incoming momentum spread of the beam results in uncaptured beam
loss of about 5\% of the injected beam, resulting in typical lost
power of 1.5 kW.  This dominant loss is readily measured with the DCCT
as shown (for smaller loss) in Fig.~\ref{AccelBeam}.

\subsection{\label{LossContColl}Loss Control Using Main Injector Collimators}

Since the slip stacking process simulation~\cite{Seiya:PAC2007}
predicted losses due to uncaptured beam, extensive measurements and
simulations of the loss process were examined to provide the
collimation system design~\cite{PAC07_CSDBLLSIFMI}. With appropriate
Booster beam emittances and Main Injector rf parameters, the loss
patterns were simulated.  The time pattern matched measurements but
the simulation suggested losses concentrated at locations with high
dispersion whereas measurements showed losses concentrated at the
Lambertson magnets in the several zero dispersion transfer regions.
By including the higher order harmonics of the Main Injector magnets,
the simulation could predict these additional loss locations.  The
simulation indicated that intercepting the uncaptured beam by defining
a limiting momentum aperture would allow a collimation system to
control the losses.  The MARS~\cite{Mokhov:1995wa,Mokhov:2007sz} energy
deposition code was used to design the secondary collimator systems to
provide adequate localization of radioactive isotope
production~\cite{Rakhno:2007zz,Rakhno:2008zz}. Using the output of the
tracking simulation as input, radiation issues were evaluated with
MARS.

In order to use the secondary collimators, local orbit displacements,
timed to impact the beam after 1\% acceleration, permitted uncaptured
beam, which had been sufficiently scattered, to strike the
collimators.  Collimators were positioned while observing aperture
requirements for other operating modes\footnote{Large displacements
were required since antiproton transfers, using the K304 kicker in the
midst of the MI300 straight section, required sufficient aperture for
the transfer orbits.}.  Measurements following an extended
commissioning phase demonstrated localization of 99\% of the
uncaptured beam loss in the collimator region~\cite{Brown:WE6RFP025}.
In addition to absorbing loss from the uncaptured beam, the secondary
collimators defined the limiting transverse aperture.  As such, the
beam removed by antidamping was preferentially lost on them.
Additional loss during the slipping process was exhibited by reduced
beam life time which was due (in part, at least) to effects of the
required large negative chromaticity. These losses also were
predominately absorbed in the collimator region.  No measurements to
separate and quantify the various secondary loss mechanisms were
devised but overall loss control showed more than 50\% of loss before
acceleration and more than 93\% of loss as acceleration began were
well contained in the collimation system.

In Fig.~\ref{AccelBeam} we show beam intensities for a typical Main
Injector cycle.  One may note the higher injected intensities for the
first and sixth injected batches which were directed toward pbar
production.  With commissioning of the collimation system and
availability of control for extraction gap losses, the 11-batch slip
stacking process became routine (2008). Intensity was limited by
activation of the injection region due to losses in the injection
kicker gaps until 2010 when the gap clearing kickers were
commissioned.  During this time and until the end of the most recent
operating period, additional intensity limits came primarily from beam
intensity and beam quality from the
Booster~\cite{Garcia:2012nf,Garcia:2012ng}. The losses at the MI8
collimators and emittance monitoring in the MI8 transfer line provided
effective monitoring of Booster beam quality.  A sequence of Booster
improvements, including a major corrector magnet upgrade, allowed
steady increases in Main Injector intensity.

\section{Aperture Improvements}

For intensity increases using slip stacking, studies and simulation
found loss mechanisms that must be met with loss control systems
(collimation, antidamping).  Within the parameters of the Main
Injector, these losses could only be localized, not eliminated.  It
was expected, however, that high intensity would emphasize losses due
to aperture limitations which could be eliminated.  A series of major
and minor efforts were applied to remove, so far as possible, the
limits due to these aperture problems.

\begin{figure}[tb]
   \includegraphics*[width=\columnwidth]{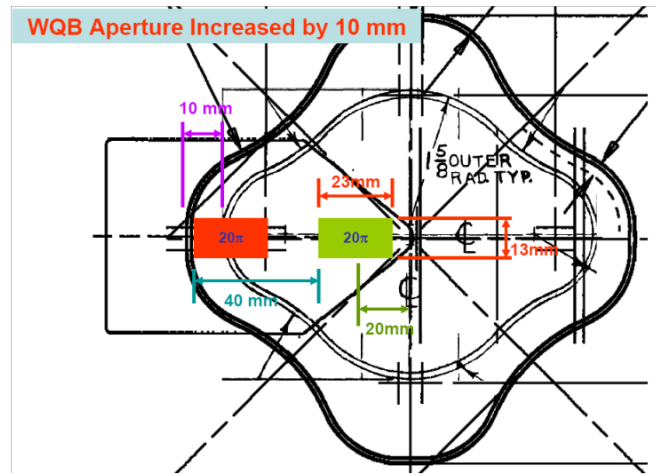}
   \caption{\label{WQBAper}Aperture Improvement using WQB large 
    aperture quadrupoles are shown in this end view of the vacuum 
    pipes (old and new) and the Lambertson which is downstream.}
\end{figure}

\subsection{\label{WideApQuad}Wide Aperture Quadrupole}

The transfer regions in Main Injector occur at straight sections which
use regular cells without dipoles.  This requires that the three
Lambertson magnets for high energy transfers~\cite{PAC95:FMI_lamb} are
split with one upstream and two downstream of the intervening
quadrupole.  The quadrupole center and the Lambertson septum are
aligned to the transverse center of the straight section.  The
circulating and transferring beams must share the quadrupole aperture,
thus placing the circulating beam at large displacement.  By
developing a set of wide aperture
quadrupoles (WQB~\cite{PAC07_WAQFMIS,PAC07_OAMILAQ}) with aperture larger by
$\sqrt{7/4}= 1.32$, improved physical aperture and much improved magnetic
field quality is available for both beams.  Fig.~\ref{WQBAper} shows
the new aperture compared with that available before the upgrade.  The
new beam pipe is illustrated by the `star-shaped' pipe surrounding
other features.  Beam apertures through the Lambertson magnets are
left (circulating) and right (transferred - usually extracted) of
center.  The beam pipe for the standard quadrupole which were used
previously in transfer regions is the smaller star shaped pipe.  The
injected beam size is shown with the new range of available positions.
WQB magnets were installed at the four high energy transfer locations
and three 8 GeV transfer locations in a 2006 facility shutdown.


\subsection{\label{PipeAlignment}Beam Pipe Alignment at Defocusing Quadrupoles}

\begin{figure}[t]
  \includegraphics*[width=\columnwidth]{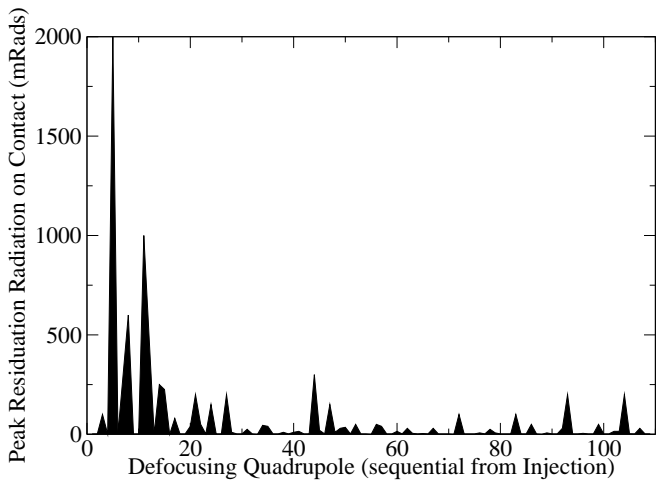}
   \caption{\label{DQuadRR}Residual Radiation on contact with top of 
    beam pipe at locations 
    between an upstream dipole and defocusing quadrupole after cooldown
    of a few hours.  Data taken on June 11, 2004.  }
    \end{figure}

\begin{figure*}[b]
   \includegraphics*[width=\textwidth]{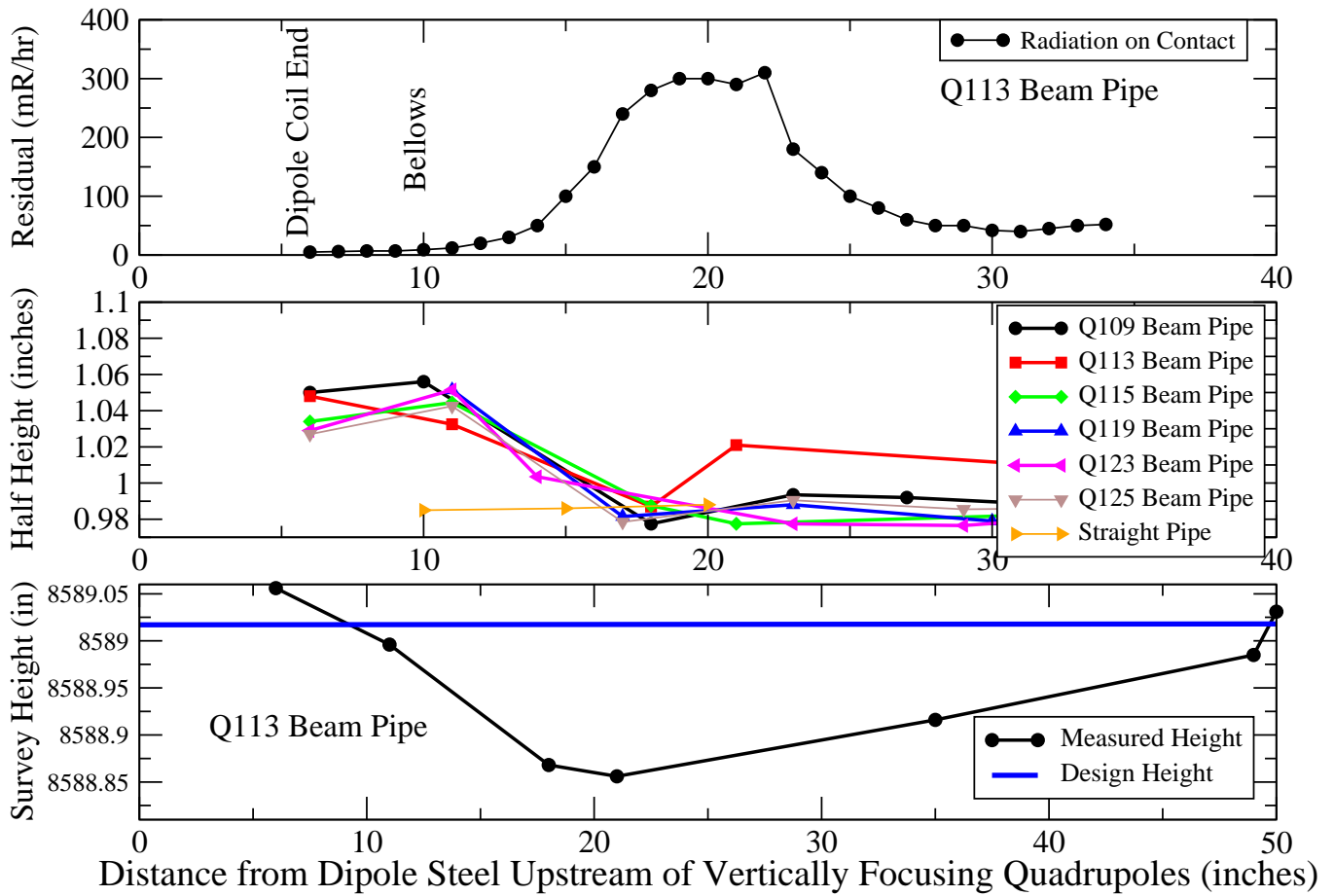}
   \caption{\label{Q113BeamPipeAp}
     Geometric distortions of beam pipes permitted beam loss 
     at locations upstream of vertically focusing quadrupoles.
     The resulting residual radiation distribution on top of beam 
     pipe near Q113 (upper panel).  Restriction due to beam 
     pipe flexing under vacuum load (center panel) for various
     locations. Restriction near Q113 from misalignment due to 
     asymmetric stress relieving of insertion-induced beam pipe stress.
     All locations experienced flexing; many experienced offset.}
\end{figure*}

The elliptical beam pipe used throughout the Main Injector (except at
transfer locations as discussed in Section~\ref{WideApQuad} and
others) provides a half aperture of about 23 mm vertically but more
than 58 mm radially.  Assuming full coupling (round beam), the similar
maximum $\beta$ values create similar beam aperture requirements:
vertically at vertically focusing quadrupoles and radially at
horizontally focusing quadrupoles.  Even adding a couple of
millimeters for momentum aperture requirements, the radial aperture is
more generous.  During the residual radiation monitoring effort
described in Section~\ref{UpgradeIMC}, a pattern of aperture
limitations was observed and understood. The pattern was significant by
creating small losses at many locations.  The initial observation is
documented in Fig.~\ref{DQuadRR}.  Localized residual radiation was
observed on the top of the beam pipe between magnets as shown in
Fig~\ref{Q113BeamPipeAp}.  This pattern was found to be due to the
flexing of the beam pipe under vacuum load.  From a point where it was
supported by a bellows (such as that shown in
Fig.~\ref{FailedBellowsQ113}), the pipe flexed to provide less
aperture by about 3 mm at a point ~0.33 m from the support or 0.5 m (18 inches)
from the upstream dipole. This beam pipe shape created losses
where, additionally, the beam pipe for many locations was displaced
because it stress-relieved after being inserted through the
quadrupole's star-shaped aperture.  For reasons not fully understood,
this stress relief motion was biased, leaving the beam pipe centerline
low by $\sim$3 mm at a fraction of the half cells.  The aperture was
reduced by 3 mm from the offset and 1.5 mm from flexing which created
the characteristic localized loss point.  Locations with severe
offsets were corrected by adding a beam pipe support and re-aligning
the pipe.  Collimation in the MI8 line was also helpful.

\subsection{Other Beam Pipe Alignment Issues}

At high beta, the 23 mm vertical aperture appeared to have several
millimeters of clearance from the beam at the three sigma beam
boundary.  Although a nominal alignment tolerance of 0.25 mm was
applied to magnetic devices, it was expected that beam pipe placement
would be adequate with only routine placement at support points.  As
we explored an unexplained loss downstream of the abort Lambertson
magnets, we discovered misalignments up to 6 mm.  Proper placement of
these beam pipes followed by application of the routine beam steering
procedures put the beams on center and greatly reduced the loss in
this area.  Comparisons of these readings in Fig.~\ref{BLM2009} and
Fig.~\ref{BLM2011} make apparent the improvement.

\subsection{\label{BIF}Bellows Installation Failure}

\begin{figure*}[htb]
   \includegraphics*[width=0.45\textwidth]{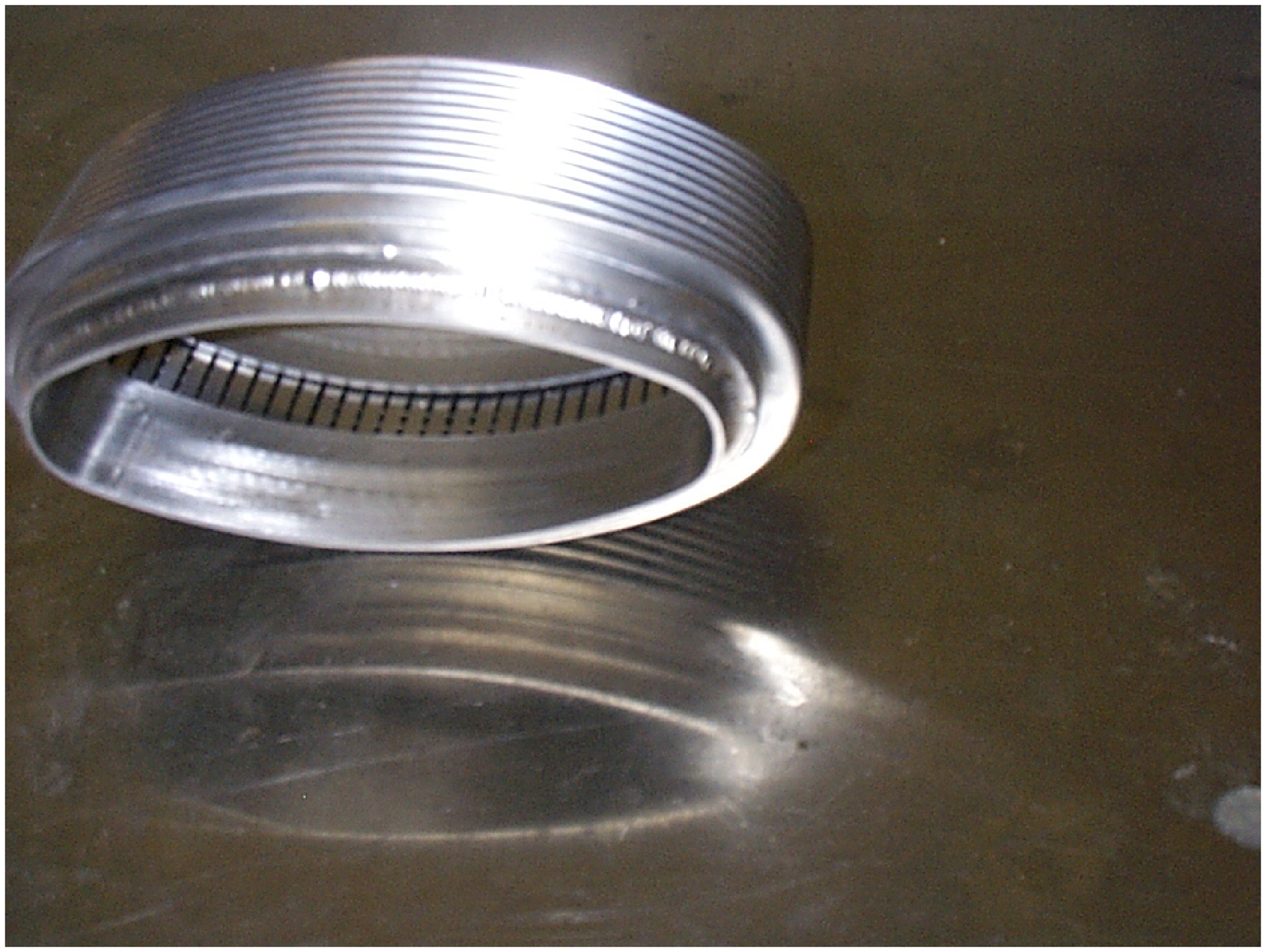}  
   \includegraphics*[width=0.45\textwidth]{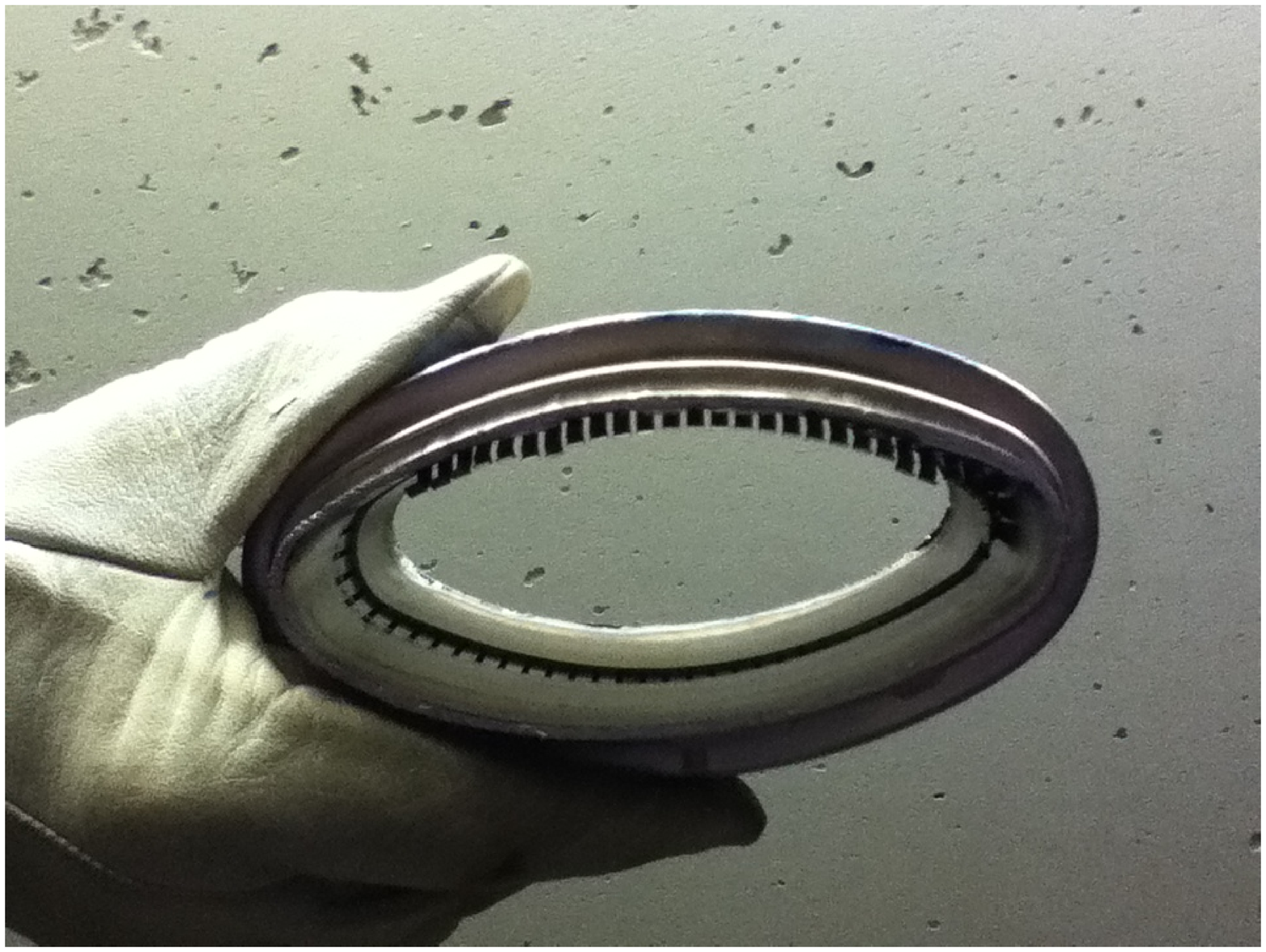}
   \caption{Elliptical formed bellows used throughout Main Injector
    to connect beam pipes.  Fingers shield beam to provide smooth
    transition for rf image currents.  Failed bellows removed on 
    March 7, 2011.  Mis-installed bellows allowed
    fingers to escape so vertical aperture was reduced by $\sim$~5 mm.
    \label{FailedBellowsQ113}}
    \end{figure*}

The vacuum system is assembled using a series of formed elliptical
bellows with rf shielding fingers to pass image currents.  A typical
bellows is shown in the left panel of Fig.~\ref{FailedBellowsQ113}.
As the loss issues around the ring were addressed by improved tuning,
a loss at the 113 loss monitor (LI113) remained.  Aperture
measurements indicated that the available vertical aperture was
reduced compared with other regular cells.  Cutting the beam pipe and
examining the space from the upstream dipole through Q113 to the
downstream dipole revealed a small limitation from beam pipe welding
and the bellows problem shown in the right panel of
Fig.~\ref{FailedBellowsQ113}.  Replacement of this bellows on March 7,
2011 removed the loss at LI113 as well as the corresponding loss
signals at LI114 and LI115. The bellows was mis-installed during a
magnet replacement on July 21, 2002 but the loss pattern change was
hard to detect until improved instrumentation was available.

\subsection{\label{LfInjTuneErr}Losses from Injection Tuning Errors}

In reviewing the various loss mechanisms, detailed analysis is limited
by complexity.  In the injection region, in addition to the beam pipe
deformations noted above, we also have identified a pattern of losses
at phase advances from the injection kicker which indicate injection
tuning problems.  At phases of 90$^o$ + n~$\times~$180$^o$ one might see
large losses when a kicker mis-fires, smaller losses when one of the
three kicker wave forms are mis-timed as well as various loss
distributions associated with slip stacked beam in the injection gap
(See Section~\ref{InjGapLoss}).  We note that the pipe alignment
issues in Section~\ref{PipeAlignment} and the bellows failure in
Section~\ref{BIF} occurred near the vertically focusing Q113.  However,
this location is also 90$^o$ + n~$\times~$360$^o$ downstream of the
injection kicker.  Additionally, problems were solved before we
acquired the current complement of instrumentation.  Losses in the
cells downstream of injection as seen in Fig.~\ref{BLM2009} and
Fig.~\ref{RR100-400-100} include many locations suggestive of
injection issues but also a number of locations where some other
problem must be responsible.  We have been unsuccessful in creating a
graphic presentation to illustrate the injection tuning issues but
have found that only a small fraction of the losses were from kicker
mis-fires.  Although we are documenting many loss issues, a variety of
features which were identified in the 2004-2006 radiation surveys were
solved without deep understanding.

\section{Other Loss Minimization Efforts}

Tuning to optimally employ the collimation, anti-damping and gap
clearing kicker improvements described above continued for a period of
several years.  The Main Injector specialists and the accelerator
operation crew employed the loss display and other tools to
progressively limit the locations where significant loss occurred.  As
a result one could note that when all systems were properly tuned, the
major losses occurred early in the cycle and were concentrated in the
collimation region and at transfer points.  We were now free to
examine a limited number of `unexplained' losses.

LAM522 and associated kickers transferred protons to the Tevatron and
accepted antiprotons from the Antiproton Source.  Despite these
complex requirements, control of the slip stacking loss combined with
careful tuning resulted in loss-free operation for extended time periods.
A campaign to tune more carefully at other transfer locations reduced losses
at each of them but some loss remained.

The improved sensitivity provided by new BPM's facilitated some
studies which required the better resolution.  In doing these
measurements, we discovered a BPM detector which had an intermittent faulty
connection.  This error had resulted in setting the orbit to wrong
position by up to 15 mm.  The large horizontal aperture of the Main
Injector allowed adequate transmission despite such errors but losses
improved when this was corrected.  Occasional other BPM failures were
quickly noted after implementation of the beam loss display.
Occasional BLM failures also allowed some additional activation before
they were identified.

\section{\label{OLC:Abort}Operational Loss Control:  Aborts and Inhibits}

The beam power of the Main Injector, especially at high energy, is
sufficient to create damage in a single pulse.  Additionally,
environmental concerns in the transport line to the NuMI target demand
very high beam quality and beam transport
control~\cite{Childress:WGF12}.  The beam abort system at MI40 can be
employed to kick all beam out in one turn.  The abort system tracks
the proton energy and sets the kickers and transport line magnets to
properly deliver the beam to the abort beam absorber.  Each beam loss
monitor channel can trigger the beam abort.  The integral loss for the
acceleration cycle and the loss in a 39 ms running sum are compared to
abort thresholds.  The abort threshold is set separately for each BLM
channel, sum type, and machine operating mode.  Special abort triggers
have been created for the NuMI operation which monitor the status of
the rf accelerating system and beam positions to protect the NuMI beam
line~\cite{Childress:WGD12}.  An abort inhibits beam for subsequent
cycles until it is reset by an operator.  Only an occasional device
failure has caused these systems to be required to protect the
facilities.

To avoid beam loss from system or device failures, the beam injection
is inhibited based on examination of the status of the beam permit.
This inhibit is applied at the upstream end of the linear accelerator.
A variety of inputs to this system inhibit further operation until
reset including status inputs from vacuum, power supply, rf, and other
accelerator systems.  As further protection, audible alarms which
require operator reset are triggered for various off-normal states.
During regular operation the most common alarm is due to the beam
energy loss (BEL) signal constructed by the BEAMS
front-end~\cite{1748-0221-6-11-T11004} which sums the beam energy loss
calculated by multiplying the incremental beam loss by the beam
energy.  The threshold for this alarm was adjusted (as improvements
permitted) to match the capability of the Main Injector and Booster
when all systems were operating well at high intensity.  These alarms
were addressed by adjusting parameters in one of the machines, by
identifying and repairing system failures, or by reducing intensity
until high quality operation could be restored.  By observing limits
on losses, we achieved higher intensities while reducing machine
component activation.

\section{Results}

We demonstrate the success in controlling and localizing loss
by examining the loss display and by reviewing the history of
residual radiation measurements.

\subsection{\label{BLMDisplay}Beam Loss Monitor Display}

As a measure of the successful loss control efforts, compare
Fig.~\ref{BLM2011} with Fig.~\ref{BLM2009}.  The injection gap losses
(top row - 8th through 10th loss monitors - green brace) have been
addressed by the gap clearing kickers.  The losses at the 17th through
19th monitors were eliminated by replacing the faulty bellows near
Q113.  The loss in the collimation region (second row - first 20 loss
monitors - cyan brace) is distributed in a more favorable way,
emphasizing the 1st and 2nd secondary collimators.  Loss at the abort
area (second row - 48th through 58th monitors - yellow brace) is
reduced.  Loss at the Recycler transfer points is eliminated (red
brace near end of first row) or greatly reduced (red brace at center
of second row).  Loss at LAM522 region is eliminated (purple brace at
end of third row -- 48th through 56th monitors) since the remaining
bar is due to a BLM pedestal offset.  In the fourth row we see loss at
LAM608 (NuMI extraction - purple brace at 12th through 22nd monitor)
and LAM620 (pbar transfer - purple brace at 38th through 43rd
monitors) are significantly reduced. Fewer signals are seen throughout
the ring while a few of the remaining bars are due to pedestal offset
in the BLM electronics.  With this fairly clean display, changes
in the loss profile provided an alert to the operations crew.

For loss reduction, the calibration of the BLM system in Rads at the
BLM detector is sufficient since we wish to eliminate any observable
loss.  For BLM calibration in protons lost, we are aware that each
loss monitor has a relation to the number of protons lost which is in
principle dependent upon the exact loss mechanism details, including
the beam orbit and to the local geometry of the machine components.
In most of the ring, we position BLM sensors on the outer wall of the
enclosure above the beam line height at the downstream end of each
quadrupole which provides a degree of uniformity for the response.
Constraining the orbit by requiring good transmission leaves little
room for a change of sensitivity.  Changes in loss are due primarily
to beam quality.  We also note that losses are nearly local but, almost
everywhere, a significant loss in one detector will also create a
response in a nearby detector.  Detailed measurements to relate lost
protons to BLM signals had mixed results and are not employed in
results for this document.  Geometric oversampling due to placing many
more BLM's at the transfer points could cause a distortion but
operationally, the losses are sufficiently concentrated that the
oversampling has little impact on overall loss evaluation.  We
provided guidance on the impact of the collimation system in
Section~\ref{LossContColl} but that assumed that all loss monitors had
the same response to a lost proton.  We believe that is a conservative
estimate since we are certain that the collimators shield the
collimation region loss monitors to make them provide smaller response
to proton loss.  Available data suggest that loss monitors have
similar calibrations in lost protons within a factor of two.

\subsection{\label{RRMonitor}Summary of Residual Radiation Monitoring}

\ifthenelse{\lengthtest{\columnwidth>100mm}}{\setlength{\FigResize}{100mm}}{\setlength{\FigResize}{\columnwidth}}

\begin{figure}[tbh]
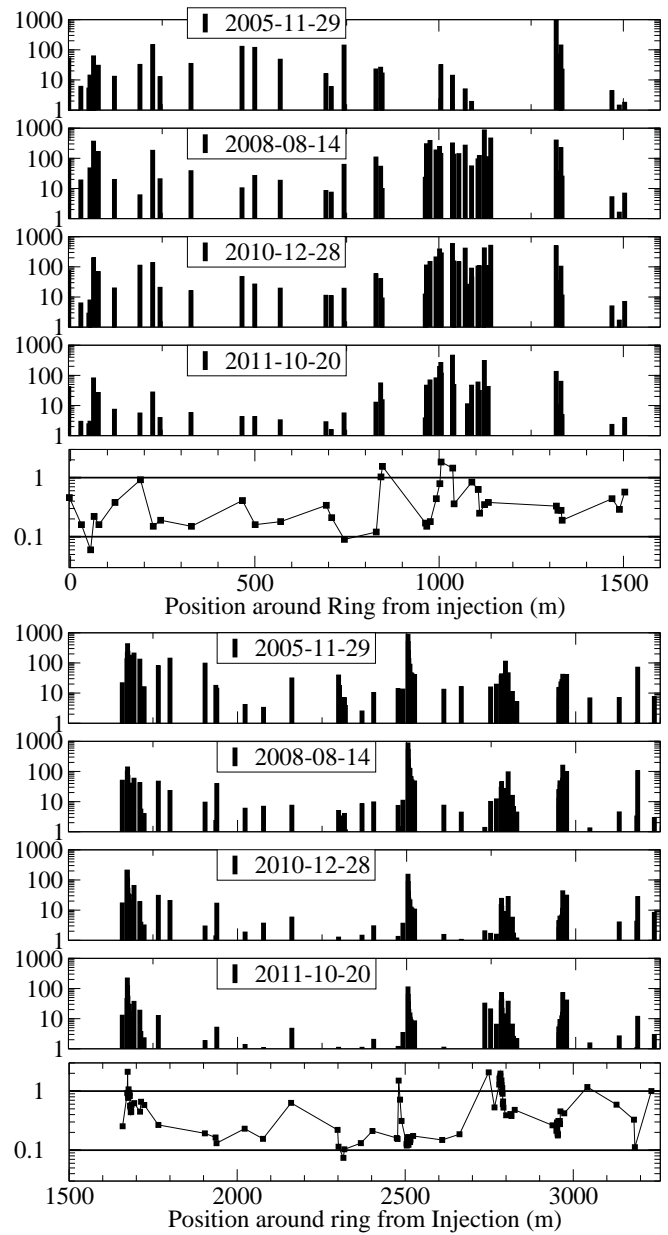

   \includegraphics*[width=\FigResize]{RR_MI100_MI400.eps}
   \includegraphics*[width=\FigResize]{RR_MI100_MI400_ratio.eps}
   \includegraphics*[width=\FigResize]{RR_MI400_MI100.eps}
   \includegraphics*[width=\FigResize]{RR_MI400_MI100_ratio.eps}
   \caption{Main Injector Residual Radiation History from Injection to
    Abort Region (upper) and from Abort Region to Injection (lower)
    measured on contact at bar-coded locations.  Ratio of measured
    radiation (November 2011/ August 2008) is shown below the
    measurement sets.  Ratio is somewhat overstated due to less
    cooldown time for 2008 data.   Major loss points: injection (50 m),
    collimation (1000 m), abort (1675 m), proton extraction
    to Tevatron (and other transfers) (2500 m), NuMI Extraction (2780 m),
    antiproton extraction to Tevatron (2950 m).
   \label{RR100-400-100}  }
\end{figure}

The definitive measure of loss control is reduced residual
radioactivity for hands-on maintenance and upgrade activities.  Losses
cannot be distributed uniformly.  This localization of the loss
implies that no single measure of radiation reduction will describe
the impact of improvements on the 3.3 km scale of the Main Injector.
The successes of the loss control campaign in the Main Injector has
lead to enormous improvements in all regions except at the
collimators.  Fig~\ref{RR100-400-100} provides snapshots of the
residual radiation at bar-coded locations selected from more than 50
such data sets.  Note the logarithmic scale where a reduction by a
factor of ten shows with a reduction of a bar by 1/3 of the vertical
scale.  We see, as we did with the loss display, the residual
radiation is greatly reduced.  Only the collimator region remains at a
nearly constant residual rate.  Detailed comparisons are best done as
discussed in Section~\ref{RRHistory} since the data were recorded
with various delays between beam activation and measurement.

\ifthenelse{\lengthtest{\columnwidth>100mm}}{\setlength{\FigResize}{95mm}}{\setlength{\FigResize}{\columnwidth}}

\begin{figure}[htb]
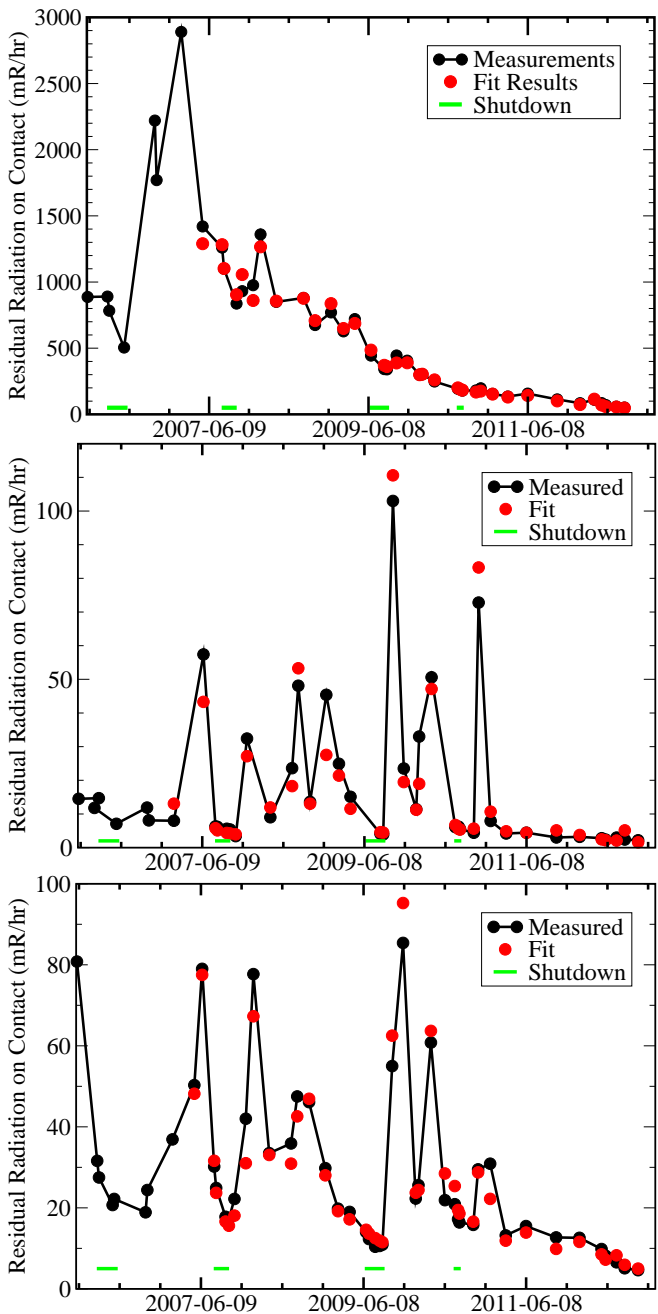

   \includegraphics*[width=\FigResize]{LAM522_RR.eps}
   \includegraphics*[width=\FigResize]{Q104DS_RR.eps}
   \includegraphics*[width=\FigResize]{S408_RR.eps}
   \caption{\label{RRhistory}Residual Radiation History at bar-coded
   locations fitted to BLM loss weighted by set of isotope half lives.
   LAM522 fit to LI522A loss (top), Top of Q104 Downstream End fitted to
   LI104 loss (center) and Top of Sextupole S408 fitted to LI408 loss
   (bottom).  Reduced loss at Q104DS followed commissioning of Gap 
   Clearing Kickers.}
\end{figure}

\subsection{\label{RRHistory}Residual Radiation History}

An alternate display of the data in Section~\ref{RRMonitor} is
available by examining the residual radiation history provided by the
measurements at bar-coded locations around the
ring~\cite{BeamsDoc3523v1}.  Selected locations are illustrated in
Fig.~\ref{RRhistory}.  The expected correlation between loss and
residual radiation has been established~\cite{Brown:TUO2C04}. Linear
fits to the correlation between half life weighted beam loss and
residual radiation history are applied using using three or four
isotopes.  A description which is adequate for most planning purposes
is achieved with three isotopes of manganese: $^{54}Mn$, $^{52}Mn$,
and $^{56}Mn$ having half life values of 312.3 days, 5.591 days, and
2.58 hours.  $^{59}Fe$ (44.5 days) or $^{51}Cr$ (27.7 days) improves
some fits.

We show measurements and fit results for the upstream end of the
Lambertson Septum LAM522 where losses during early (5+2) slip stack
operation resulted in very high radiation levels. Improved tuning
followed by implementation of the antidamping for the extraction gaps
reduced the loss so the residual levels began to fall.  Later we
achieved loss-free operation at this location.

The losses from the injection gap impacted devices in several half
cell locations downstream from the injection kicker (K103). We
illustrate this with the history at Q104 Downstream.  Losses were
monitored and beam intensity was limited to keep the residual
radiation at a level suitable to permit tunnel modifications during
the 2009 facility shutdown.  At this time, the GCK magnets were
installed but the power supplies and cables awaited additional tunnel
time.  The peak in radiation following the 2009 shutdown was the
result of relaxed requirements on beam loss monitor values.  GCK
commissioning followed the 2010 shutdown and after successful
commissioning, the loss in this region dropped to small values.

As another example, we show the loss history at the S408 sextupole.
Detailed measurements at this location (and several others) revealed
radial loss.  Examination of the BLM signals frequently showed loss
at transition although other loss times contributed.  This was a 
situation where the Beam Loss Monitor Display proved very helpful.
Significant radiation issues can build up with only a small impact 
on transmission.  As other problems were addressed, this loss point 
received appropriate attention and losses were mitigated.

\section{Summary of High Intensity Operation}

The beam properties achieved for high intensity operation of the Main
Injector matched the goals set in the Proton
Plan~\cite{ProtonPlan}. They are summarized in
Table~\ref{Table:BeamProp}.  When operating to maximize both neutrino
and pbar production, we employed the 9+2 slip stacking mode described
above (mixed mode) at the 2.2 second repetition rate achievable.  The
high energy physics program required additional operating modes.  The
Main Injector (as its name suggests) is the source of 150 GeV protons
and antiprotons for the Tevatron Collider.  The transfer of 150 GeV
beam and setup for that required interruptions to the mixed mode
operation.  More frequently the process was interrupted to transfer
antiprotons from the Accumulator to the Recycler.  Both of these
processes were gradually optimized to improve high intensity
productivity.  A portion of the Main Injector time was devoted to slow
spill operation which provided 4.05 seconds of extraction at 120 GeV
for the test beam and fixed target experimental program.  When the
NuMI beamline was unavailable to take beam, a pbar-only mode with two
slip stacked batches was operated.  Since the collection and cooling
power of the Antiproton Source was saturated by the standard 2.2
second repetition rate, the pbar-only mode repetition rate was not
maximized.  During the brief intervals when the Tevatron or Antiproton
Source were unavailable, a NuMI-only mode with 11 batch slip stacking
was employed.  At the end of the Tevatron run, a 9 batch slip stacking
mode with 2.066 second cycle time was created to maximize beam power
while observing a per pulse intensity limit of $3.75\times10^{13}$
protons per pulse designed to protect the neutrino production target
from thermal shocks.  

\begin{figure}[htb]
   \includegraphics*[width=\columnwidth]{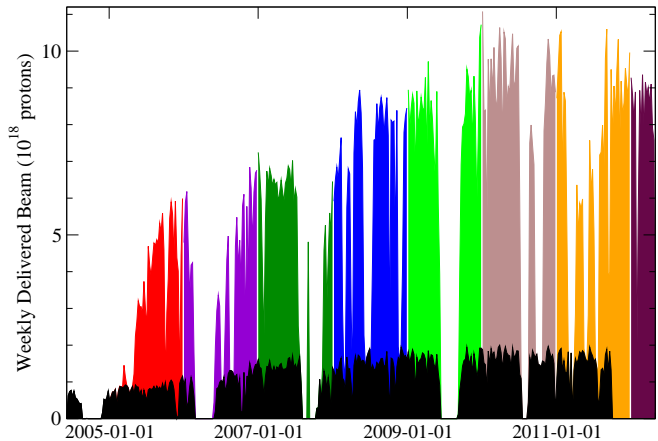}
   \caption{\label{WSumP}Weekly Summary of Protons Accelerated in Main
      Injector for delivery to the pbar and NuMI targets.  This
      stacked bar chart shows pbar weekly beam in black bars with
      colored bars indicating the NuMI weekly beam (each year a new
      color) so the top of the bar indicates the total weekly
      delivered 120 GeV beam.}
\end{figure}

In Fig.~\ref{WSumP} we document the weekly sum of protons accelerated
to 120 GeV from 2004 through 2012.  With a typical intensity of
$4.2\times10^{13}$ protons per cycle, the maximum weekly sum is
1.15$\times10^{19}$ protons/week. We find that by the beginning of
2008, we were achieving up to 50\% of the this rate but by
commissioning the 11 batch slip stacking and the collimation, we
achieved 70\% of this mark during 16 weeks of that year.  Steady
progress in 2009 was culminated with two weeks which achieved 90\% of
that target.  In 2010 there were 8 weeks above 88\% of that target.
That year we achieved our peak weekly integrated beam of
$1.109\times10^{19}$ protons per week or 96.4\% of this goal.

These beam power limitations for the Main Injector are principally set
by the capabilities of the Fermilab Booster in combination with design
properties of the Main Injector.  The momentum aperture of the Main
Injector is adequate for slip stacking injection.  The Booster fifteen
hertz structure sets a requirement for slipping speed and thereby for
frequency difference for the two slipstacking rf systems.  The
injection buckets created by 100 kV rf systems have as large an
acceptance as possible without overlapping the buckets.  The
longitudinal emittance and especially the (related) momentum spread of
the Booster beam does not match these buckets at the desired
intensity.  The uncaptured beam loss in high intensity operation is
due to Booster beam with too large $\delta p/p$.  Additional losses in
the injection and extraction gaps are also dictated by Booster
emittance in combination with bunch-to-bunch phase offsets due to
coupled bunch instabilities in the Booster. Additional losses due to
large negative chromaticity operation are observed but are small.
These also would likely be smaller with lower Booster transverse
emittance and would not be a problem if slip stacking injection were
not required.  In summary, using the current equipment, the intensity
capability of the Main Injector is not challenged until a more intense
injector is available\footnote{A limit due to rf capability is
expected at intensities about 30\% higher than the current per pulse
operation~\cite{BeamsDoc1927}.}.

\section{Conclusions and Observations}
 
The requirement to produce abundant antiprotons and neutrinos to match
Fermilab's High Energy Physics program required enhancements to the
initial Main Injector configuration.  The Proton Plan~\cite{ProtonPlan}
as developed in 2004 - 2006 identified and addressed the limitations
in the Main Injector.  Improvements to the Fermilab Booster were also
addressed.  The results included improved beam properties as well as
lower losses in the Booster.  The performance envelope of the Booster 
continues to define the intensity limit for Main Injector operation.

Main Injector loss control efforts have been directed at maintaining
low residual radiation for maintenance and upgrade activities.  This
has focused efforts on optimizing the use of the Main Injector
Collimation system and in vigilant attention to removing localized
loss points since even fairly small loss will allow accumulation of
residual radiation which can impact planning for repairs.

The 2012-2013 facility shutdown will implement a series of
modifications designed to permit operation of the Main Injector at 700
kW beam power at 120 GeV~\cite{Derwent:2012zz}.  This will be achieved
with modest improvements in per cycle beam intensity and by enhancing the
repetition rate from 2.2 seconds to 1.3 seconds by employing the
Recycler Ring as an 8 GeV stacking ring.  12 Batch slip stacking in
the Recycler will be followed by recapture in the 1-MV Main Injector
rf system.  When Recycler modifications and Booster repetition rate
enhancements~\cite{Garcia:2012nf,Garcia:2012ng} are complete we expect
operation at 700 kW to be available. We are expecting radiation issues
to increase only proportional to the beam power

\begin{acknowledgments}

The high intensity operation of the Main Injector was accomplished by
the efforts of the entire Fermilab Accelerator Division. We thank the
many helpful people in the other machine departments and support
departments and the Operations Department for their on-going
commissioning and tuning efforts.  The Fermilab Technical Division
created many devices and provided support for maintenance and upgrade
activities.  We thank the Fermilab experimenters for their support and
encouragement.

\end{acknowledgments}

\bibliography{MIHiIntNLoss_prstab}

\end{document}